\documentstyle[12pt]{article}
\begin{document}
%  Greek letters
\def\a{\alpha}
\def\b{\beta}
\def\ch{\chi}
\def\d{\delta}
\def\e{\epsilon}
\def\f{\phi}
\def\g{\gamma}
\def\h{\eta}
\def\i{\iota}
\def\j{\psi}
\def\k{\kappa}
\def\l{\lambda}
\def\m{\mu}
\def\n{\nu}
\def\o{\omega}
\def\p{\pi}
\def\q{\theta}
\def\r{\rho}
\def\s{\sigma}
\def\t{\tau}
\def\u{\upsilon}
\def\x{\xi}
\def\z{\zeta}
\def\D{\Delta}
\def\F{\Phi}
\def\G{\Gamma}
\def\J{\Psi}
\def\L{\Lambda}
\def\O{\Omega}
\def\P{\Pi}
\def\S{\Sigma}
\def\U{\Upsilon}
\def\X{\Xi}
\def\T{\Theta}
\def\pp{\partial }
\def\pb{\bar{\partial}}
\def\be{\begin{equation}}
\def\ee{\end{equation}}
\def\ben{\begin{eqnarray}}
\def\een{\end{eqnarray}}

\addtolength{\topmargin}{-0.8in}
\addtolength{\textheight}{1in}
\hsize=16.5truecm
\hoffset=-.5in
\baselineskip=7mm

\thispagestyle{empty}
\begin{flushright} \ March \ 1996\\  SNUTP 96-016/  hep-th/9603027\\
\end{flushright}

\begin{center}
 {\large\bf Gravitating BPS Dyons witout a Dilaton
 }\\[.1in]
\vglue .5in
Choonkyu Lee 
\vglue .3in
{\it Department of Physics and Center for Theoretical Physics
\\
Seoul National University. Seoul, 151-742, Korea } 
\vglue .2in
and
\vglue .2in   
Q-Han Park\footnote{ E-mail address; qpark@nms.kyunghee.ac.kr }
\vglue .3in
{\it Department of Physics, Kyunghee University\\
Seoul, 130-701, Korea}
\\[.3in]
{\bf ABSTRACT}\\[.2in]
\end{center}
\vglue .2in
We describe curved-space BPS dyon solutions, the ADM mass of which saturates 
the gravitational version of the Bogomol'nyi bound. This generalizes 
self-gravitating BPS monopole solutions of Gibbons et al. when there is no 
dilaton.

\newpage
Recently, Gibbons et al. [1] showed that the well-known BPS monopole 
equations [2] of the flat-space Yang-Mills/Higgs theories find natural 
curved-space counterparts in the context of certain supergravity-motivated 
theories. A crucial feature in these theories is that, in the presense of the 
additional attractive force due to  gravity (and possibly due to the dilatonic interaction), the force balance between equal-sign monopoles is maintained 
by including an additional Abelian vector potential $A_{\m }$ (the graviphoton)  which has a nonrenormalizable coupling to the Yang-Mills magnetic charge density. It has also been noted in [1] that these static monopole configurations 
in fact saturate the gravitational version of the Bogomol'nyi bound, derived by  
Gibbons and Hull [3] some time ago. In the notation of [1], the latter bound reads
\be
M \ge |Q|,
\ee
where $M$ is the ADM mass and $Q$ is the total electric charge with respect to the graviphoton field. 

To study the physical character of these solutions, it should be important to have 
related self-gravitating dyon solutions.
In flat space-time the very system allowing BPS monopoles is known to admit 
also dyon solutions which saturate the appropriate Bogomol'nyi bound[4]. 
Including gravity, however, the situation appears to be different and in 
this paper we report our study on this issue. We will here restrict our attention to gravitating BPS dyon solutions in the model without a dilaton, viz., our SU(2) Yang-Mills/Higgs matter fields will be coupled only to gravity and a graviphoton field. Note that the system without a dilaton is described by a 
simpler Lagrangian than that with a dilaton; yet, it does not correspond to a certain limiting 
case of the latter[1].
For discussions on BPS dyons in a model including a dilaton and some other fields, see a very recent paper by Gibbons and Townsend[5]. 
 
The model chosen in [1] for curved-space BPS monopoles is based on the action
\be
S^{'}_{\mbox{full}} = S_{\mbox{gravity}} + S^{'}_{\mbox{matter}}
\ee
with
\ben
S_{\mbox{gravity}} &=& { 1 \over 16\pi G}\int d^{4}x \sqrt{-g} \{ 
R - F^{\m\n }F_{\m\n } \} \ , \ ( F_{\m\n } = \pp_{\m }A_{\n} - \pp_{\n }
A_{\m } ) \\ 
S^{'}_{\mbox{matter}} &=& {1 \over e^{2}}\int d^{4}x \mbox{ tr }\{ {1\over 2}
\sqrt{-g } (G_{\m\n }G^{\m\n } ) + \sqrt{-g } (D_{\m }
\Phi D^{\m }\Phi ) - {1 \over 2} \e^{\m\n\l\d }F_{\m\n }
(\Phi G_{\l\d })\},
\een
where $G_{\m\n } = \pp_{\m }B_{\n } - \pp_{\n }B_{\m } + [ B_{\m } \ , \ B_{\n }]$ is the field strength for an SU(2) Yang-Mills vector potential $B_{\m } = 
B_{\m }^{a}T^{a} , \ \Phi = \Phi^{a}T^{a}$ is the Higgs field in the adjoint 
representation, and $D_{\m }\Phi = \pp_{\m }\Phi + [ B_{\m } \ , \ \Phi ]$.
(We assume that $[T^{a} ,  T^{b}] = \e_{abc}T^{c} \ , \ T^{a \dagger }= 
-T^{a}$ and tr$(T^{a}T^{b}) = -{1 \over 2}\d^{ab}$). 
In order to have the Gibbons-Hull bound (1) to be saturated, 
the metric and graviphoton field may be chosen as[1,5]
\be
ds^{2} = -e^{-2\phi }dt^{2} + e^{2\phi }dx^{i}dx^{i} ,~~ A_{0} = \pm e^{-\phi }
\ , ~~ A_{i} = 0,
\ee
where $\phi = \phi(\vec{x})$ is a function of spatial coordinates only. 
Because of the Einstein and the graviphoton field equations, the choices (5) 
in turn imply
\be
T^{\mbox{matter }}_{ij} = T^{\mbox{matter}}_{0i} = J_{i} = 0, \ (i,j = 1,2,3),
\ee
where $T_{\m\n}^{\mbox{matter}}$ denotes the matter stress energy tensor 
and $J_{\m }$ the graviphoton current. The conditions (6) are 
realized only when the matter action (4) by itself happens to define the 
Bogomol'nyi or self-dual system (in the presence of the background 
metric and the graviphoton fields (5)). That is indeed the case thanks to the 
last term in (4) (which was motivated by supersymmetry [1]). 
The corresponding Bogomol'nyi equations read
\be
G_{ij} = \pm e^{\phi }\e^{ijk}D_{k}\Phi \ , ~~~ B_{0} = 0,   
\ee
which generalize the flat space BPS monopole equations in an obvious way.
Then, we have one more equation coming from the Einstein (or graviphoton)
equations
\be
e^{-\phi }\nabla^{2} e^{\phi } = {8\pi G \over e^{2}}\mbox{tr} (D_{i}\Phi 
D_{i}\Phi ).
\ee
For the existence of soliton solutions to these coupled equations with 
non-zero `matter' magnetic charge, see the appendix of [1].

But it must be noted that the matter action (4), with the background metric and the 
graviphoton fields specified as in (5), corresponds to a self-dual system 
only for solitons with vanishing matter electric charge. So certain 
extension of the theory must be made in order to incorporate BPS dyons 
in the above scheme. Using N=2 supersymmetry as a guide [6], a rather obvious 
step will be to introduce an additional Higgs field $S=S^{a}T^{a}$ (to 
complete an N=2 vector multiplet). 
(Then we shall see that the vacuum for dyon solutions is in fact different from the 
vacuum for neutral monopoles.)
To be more definite, we will below look for an appropriate curved-space generalization of the bosonic part of the flat-space
N=2 super Yang-Mills theory, described by the action [7]
\be
S_{\mbox{flat}} = {1 \over e^{2}}\int d^{4}x \{ \mbox{tr} ({1\over 2}G_{\m\n }G^{\m\n})
+ \mbox{tr}(D_{\m }\Phi D^{\m }\Phi ) + \mbox{tr}(D_{\m }SD^{\m }S )
+ \mbox{tr} ([\Phi , S][\Phi , S])\} .
\ee
(The last commutator square term is actually not important for our purpose, 
since we find $[\Phi , S]=0$ at least for our classical solutions to be discussed). 
For the discussion of the Bogomol'nyi bound and dyon solutions in the 
theory (9), see [6,7]. We here add just one comment - this theory possesses the global chiral symmetry
\be
\Phi \rightarrow \Phi^{'} = \cos{\b }~ \Phi - \sin{\b }~S \ , \ 
S \rightarrow S^{'} = \sin{\b } ~\Phi + \cos{\b }~ S ,
\ee 
which plays some role as regards the dual symmetry involving BPS dyons. 

Our immediate problem is to find an appropriate self-dual generalization of the theory (9) in the curved background specified by (5). Denoting such action by $S_{
\mbox{matter}}$, we can then base our model for curved-space matter dyons on 
the action 
\be
S_{\mbox{full}} = S_{\mbox{gravity}} + S_{\mbox{matter}} ,
\ee
where $S_{\mbox{gravity}}$ is given by (3). According to our investigations, 
the requirement of self-duality for $S_{\mbox{matter}}$ demands that we choose 
the form
\ben
S_{\mbox{matter}} &=& {1\over e^{2}}\int d^{4}x\sqrt{- g}\{ 
{1\over 2} \mbox{tr}(G_{\m\n} G^{\m\n}) + \mbox{tr} 
(D_{\m }\Phi D^{\m}\Phi ) + \mbox{tr}
(D_{\m }SD^{\m }S ) + \mbox{tr}([\Phi , S][\Phi ,S]) 
\nonumber \\ && - \mbox{tr}[(\mbox{}^{*}F
^{\m\n}\Phi + F^{\m\n}S )G_{\m\n} - (\alpha \mbox{}^{*}F^{\m\n}\Phi 
+ {1\over 2}F_{\m\n}S) F_{\m\n }S  ]\},
\een
where $\alpha $ is a constant to be fixed later and 
$\mbox{}^{*}F^{\m\n} = {1\over 2\sqrt{-g}}\e^{\m\n\l\d }F_{\l\d }$. This matter action reduces to $S^{'}_{\mbox{matter}}$(see (4)) if the scalar field
$S$ is taken to be zero. One might be surprised by the fact that the last term in (12) is not invariant under the chiral transformation (10), 
but this cannot be avoided. From the matter action (12) we find the matter stress energy tensor (to be identified as the gravitational source)
\ben
T^{\mbox{matter}}_{\m\n } &=& -{2\over e^{2}}\mbox{tr} (G^{\m }_{ \ \l }G^{\n \l }- {1\over 4}g^{\m\n }G_{\l\d }G^{\l\d }) - {2 \over e^{2}}\mbox{tr}(
D^{\m }\Phi D^{\n }\Phi - {1 \over 2}g^{\m\n }D_{\l }\Phi D^{\l }\Phi )
\nonumber \\ &&
-{1 \over e^{2}}g^{\m\n }F^{\l\d}\mbox{tr}S(G_{\l\d } - {1\over 2}F_{\l\d }S) 
 + {2 \over e^{2}}F^{\m }_{\ \l}\mbox{tr} S(G^{\n\l }-{1\over 2}F^{\n\l }S )
\nonumber \\ && + {2\over e^{2}}F^{\n }_{\ \l }\mbox{tr} S(G^{\m\l } 
 - {1 \over 2}F^{\m\l }S)
\een   
and the graviphoton current (to be equated with ${1 \over 4\pi G }F^{\m\n }_{ \ \ ;\m })$
\be
J^{\n } = -{2\over e^{2}}\mbox{tr}(D_{\m }\Phi )^{*}G^{\m\n } - {2 \over e^{2}} [\mbox{tr}(SG^{\m\n })]_{;\m } + {4\alpha  \over e^{2}}\mbox{}^{*}F^{\m\n} [
\mbox{tr}(\Phi S)]_{,\m } + {2 \over e^{2}}[F^{\m\n }\mbox{tr} S^{2}]_{;\m }.
\ee 

Let us now derive the BPS dyon equations for our system. From the Hamiltonian 
for the theory (12), the (static) matter energy is found to be    
\ben
\cal{E} &=& {1 \over e^{2}}\int d^{3}x \{ - {1 \over 2}\sqrt{-g}\mbox{tr}(G_{ij}
G^{ij}) - \sqrt{-g}\mbox{tr}(G_{0i}G^{0i}) - \sqrt{-g}\mbox{tr}(D_{i}\Phi 
D^{i}\Phi ) \nonumber \\ && + \sqrt{-g}(D_{0}\Phi D^{0}\Phi ) - 
\sqrt{-g}(D_{i}SD^{i}S ) + \sqrt{-g}\mbox{tr}(D_{0}SD^{0}S ) - 
\sqrt{-g}\mbox{tr}
([\Phi , S][\Phi 
, S]) \nonumber \\
&& + \mbox{tr}(\mbox{}^{*}F^{ij}\Phi + F^{ij}S )G_{ij}
 - \alpha \mbox{}^{*}F^{\m\n }F_{\m\n}\mbox{tr}(\Phi S) 
- {1\over 2}
F^{\m\n }F_{\m\n}\mbox{tr}S^{2}\} .
\een
But from what has been assumed in (5),
\ben    
-g^{00} &=& e^{2\phi } , ~~ g_{ij} = e^{2\phi }\d_{ij} ,~~ \sqrt{-g}=e^{2\phi }   
\nonumber \\ 
F_{0i} &=& \mp \pp_{i}e^{-\phi } , ~~ \mbox{}^{*}F_{ij} = \pm \e^{ijk}\pp_{k}
e^{\phi } ,~~ F_{ij} = \mbox{}^{*}F_{0i} =  0 .
\een
Inserting these ansatze into (15), the matter energy can be expressed as
\ben
\cal{E} &=& -{1\over 2e^{2}}\int d^{3}x\{ e^{-2\phi }\mbox{tr}(G_{ij} \mp
 e^{\phi }\e^{ijk}D_{k}\Phi )^{2} + 2e^{2\phi }\mbox{tr}(G_{0i} 
\pm e^{-4\phi }D_{i}S )^{2} + 2e^{4\phi }\mbox{tr}(D_{0}\Phi D_{0}\Phi )
\nonumber \\
&&+ 2e^{4\phi }\mbox{tr}([\Phi , S][\Phi , S]) 
+ 2e^{4\phi }\mbox{tr}
(D_{0}S D_{0}S) \pm 2e^{-\phi }\e^{ijk}\mbox{tr}(G_{ij}D_{k}\Phi )  
\mp 4e^{\phi }\mbox{tr}(G^{i0}D_{i}S)
\nonumber \\ && \pm 2\e^{ijk}(\pp_{i}e^{-\phi })
\mbox{tr}(\Phi G_{jk}) - 2\nabla\phi \cdot \nabla\phi \mbox{tr} S^{2}\}.
\een
At this point, it must be noted that the field $B_{0}$ assumes the role
 of a Lagrange multiplier, which puts our matter fields under the 
Gauss constraint
\ben
0 &=& {\d S_{\mbox{matter}} \over \d B_{0}(x) } \nonumber \\
&=& e^{-2\phi  }D_{i}(e^{2\phi }G^{i0}) - [\Phi , D^{0}\Phi ]
-[S, D^{0}S] \pm \pp_{k}(e^{-\phi })D_{k}S \mp e^{-2\phi }
(\nabla^{2 }e^{\phi} )S.
\een
Making use of (18) and the Bianchi identity $\e^{ijk}D_{i}G_{jk} = 0
$, we can then cast (after some straightforward algebra) the above matter 
energy as
\ben
{\cal{E}} &=& {1 \over 2e^{2}}\int d^{3}x [ -e^{-2\phi }\mbox{tr} (G_{ij} \mp 
e^{\phi }\e^{ijk}D_{k}\Phi )^{2} - 2e^{2\phi }\mbox{tr} 
\{ G_{0i} \pm e^{-\phi }D_{i}S \pm (\pp_{i}e^{-\phi })S\}^{2} 
\nonumber \\ && -2\mbox{tr}(D^{0}\Phi \mp e^{\phi }[\Phi 
, S])^{2} 
- 2e^{4\phi }\mbox{tr} (D_{0}S)^{2} \nonumber \\
&& \mp 2\pp_{i}\{ e^{-\phi }\e^{ijk}\mbox{tr}
(\Phi G_{jk} ) 
- 2e^{\phi }\mbox{tr} (S[G^{i0} \pm (\pp_{i}e^{-\phi })S])\}]. 
\een 
This has the desired structure, the integrand consisting of the sum of squared terms up to the total derivative term.

For any finite energy configuration in the above theory, the scalar fields $\Phi^{a}$ and $S^{a}$ may be assumed to have the asymptotic forms
\be
 \ \Phi^{a} \rightarrow \Phi_{\infty }\hat{\Phi}^{a}
\ , \ S^{a} \rightarrow S_{\infty }\hat{\Phi }^{a} ~~ \mbox{as} ~~
r \rightarrow \infty  ,
\ee
where $\hat{\Phi }^{a}\hat{\Phi }^{a} =1.$ The total `matter' magnetic and electric charges are now specified by
\be
g_{m} = \oint_{ r = \infty }d^{2}S^{i}\e^{ijk} \mbox{tr} (\hat{\Phi }G_{jk} )
\ , \ q_{m} = \oint_{r =\infty }d^{2}S^{i} 2 \mbox{tr} (\hat{\Phi }G^{0i} ).
\ee
On the other hand, we define the graviphoton electric charge $Q$ as
\be
Q = -{1 \over 4\pi G }\int d^{2}S^{i} F^{0i} .
\ee
This leads to the formula $Q = \int d^{3}x \sqrt{- g}J^{0}$ upon using the 
graviphoton field equations). Then, based on (19), we secure the Bogomol'nyi 
bound (note that $\phi = O(1/r)$ as $r \rightarrow \infty)$
\be
{\cal{E}} \ge {1 \over e^{2}}|\Phi_{\infty }g_{m} + S_{\infty }(q_{m} - 
4\pi GS_{\infty }Q )|.
\ee
We here note that a simpler form of the bound can be given if the field 
configurations satisfy the graviphoton field equations. It is because 
the relation
\be
Q = {1 \over e^{2} + 4\pi GS_{\infty }^{2}} (\Phi_{\infty }g_{m} 
+ S_{\infty }q_{m}),
\ee
then holds as a simple consequence of (14) and our definitions for 
various charges. Using (24), the bound (23) can alternatively be written as
\be
{\cal{E}} \ge {1 \over e^{2} + 4\pi G S^{2}_{\infty }}|\Phi_{\infty }g_{m} 
+ S_{\infty }q_{m }| , 
\ee
or, more succintly, as ${\cal E} \ge |Q|$ (in agreement with (1)).
Also evident from (19) is that this Bogomol'nyi bound is saturated 
if and only if the matter fields satisfy the curved-space BPS equations
\ben
G_{ij} &=& \pm e^{\phi }\e^{ijk}D_{k}\Phi ,\nonumber \\
G_{0i} &=& \mp D_{i}(e^{-\phi }S ) ,\nonumber \\
D_{0}S &=& 0 ,\nonumber \\
D^{0}\Phi &=& \pm e^{\phi }[\Phi , S] .
\een
The last three equaions in Eq.(26) are solved by 
\be
B_{0} = \pm e^{-\phi }S .
\ee
Also, from our experience in the flat-space theory case [6,7], 
we may here assume that $S(\vec{x})$ and $\Phi (\vec{x})$ are 
proportional to each other,viz.,
\be
S(\vec{x}) = \tan {\g }~ \Phi (\vec{x}), \ \   (\tan {\g } = 
S_{\infty }/\Phi_{\infty } )
\ee
or, equivalently, we write 
\be
\Phi(\vec{x}) = \cos{\g } ~H(\vec{x}), \ \   S(\vec{x}) = \sin{\g }~ H(\vec{x}).
\ee
Inserting (29) into (26) then yields the equations analogous to the flat-space BPS equations of [4]:
\ben
G_{ij} &=& \pm \cos{\g }~e^{\phi }\e^{ijk}D_{k}H , \nonumber \\
G_{0i} &=& \mp \sin{\g }~D_{i}(e^{-\phi }H).
\een  
Note that, given values of $H_{\infty }(=\sqrt{\Phi_{\infty }^{2} + 
S_{\infty }^{2}})$ and $\g $, the charges $q_{m}$ and $Q$ of the corresponding 
dyon solution (satisfying the graviphoton field equations also) are
 determined as
\be
q_{m} = g_{m} \tan{\g }~(1+ {4\pi G \over e^{2}}H_{\infty }^{2}), ~~~~~~  
Q= g_{m}{H_{\infty } \over e^{2}\cos{\g }}.
\ee

Readers may verify that our dyon fields, determined by (26)-(28), satisfy
 the Gauss constraint (18). By the standard argument, they also solve other matter field equations. To check the consistency with the Einstein and graviphoton field equations, we must look into the stress energy tensor and the graviphoton current. 
Using the above results with the formulae (13) and (14), we find that 
the latter quantities are given by (cf.(6))
\ben
T^{\mbox{matter}}_{\m\n} &=& \d_{\m 0}\d_{\n 0}({-2\over e^{2}})e^{-4\phi }\mbox{tr}(D_{i}HD_{i}H), \nonumber \\
J^{\n } &=& \d_{\n 0}(\pm {2\over e^{2}})e^{-\phi }\mbox{tr}(D_{i}HD_{i}H ) 
\een
if we fix the constant $\alpha = 1/2$ in Eqs.(12) and (14).
Consequently, all Einstein and graviphoton field equations can be fulfilled if 
the function $\phi (\vec{x})$ satisfies (cf. (8))
\be
e^{-\phi }\nabla^{2}e^{\phi } = {8\pi G \over e^{2}}\mbox{tr} (D_{i}HD_{i}H ).
\ee
We here note that the solution to (30) and (33), denoted as $(B_{i}(\vec{x})
, H(\vec{x}), \phi (\vec{x}))$, can immediately be given once the functions $
(\tilde{B}(\vec{x}), \tilde{H}(\vec{x}), \tilde{\phi}(\vec{x}))$, 
which solve the same equations but with $\g = 0$(i.e., the BPS monopole 
case discussed in [1]), are known.  This is done by the simple substitution
\ben
B_{i}(\vec{x}) &=& \cos{\g }~\tilde{B}_{i}(\vec{y})|_{\vec{y} = \vec{x}\cos {\g }} \nonumber \\
H(\vec{x}) &=& \tilde{H}(\vec{y})|_{\vec{y} = \vec{x}\cos{\g } }, ~~
\phi (\vec{x}) = \tilde{\phi }(\vec{y})|_{\vec{y} = \vec{x}\cos{\g }}.
\een
A corollary of this observation is the existence of rotationally symmetric 
dyon solutions, based on the analysis of [1]. Possible global issues 
concerning these solutions are under investigation. 

A few remarks are in order. Our curved-space BPS dyon solution, being 
possible only when $S_{\infty }\ne 0$, is not a solution in the same vacuum 
as that of the neutral monopole case (for which we have $S(\vec{x})=0$).
This is in a marked contrast to the case of the flat-space self-dual 
system defined by the action (9), and the absence of the global chiral symmetry (see (10)) in our matter action (12) is responsible for this peculiar 
situation found in the curved-space case. (A similar phenomenon has also 
been observed in the dyon model considered in [5]). Also, it is desirable to 
clarify a possible connection between our curved-space self-dual action and certain 
version of supergravity models[8,9]. A typical term of supergravity model which might be relevant to 
the action (12) is the interaction term involving the two-index tensor field $Y_{\alpha 
\beta }$ and the $SU(2)$ gauge fields $B_{\alpha }^{a}T^{a}$ in the six-dimensional 
gauged N=4 supergravity[9],
\be
L_{I} ={1 \over 8} \e ^{\a \b \g \d \l \s }Y_{\a\b }\mbox{ tr }(G_{\g \d }G_{\l 
\s }).
\ee
Consider a dimensional reduction by setting
\ben
(ds_{6})^{2} &=& (dx^{6})^{2} + (dx^{5} - 2K )^{2} + (ds_{4})^{2} 
\nonumber \\
Y &=& Y_{5\m} dx^{5} \wedge dx^{\m } = A_{\m }dx^{\m } ~~(\m = 0,1,2,3 )
 \nonumber \\
B &=& B_{\m }dx^{\m } + S(dx^{5} - 2K ) + \Phi dx^{6} \nonumber \\
K &=& A_{\m }dx^{\m }
\een
where $K$ is a one-form of four-dimensional spacetime and $\Phi , S$ are four-dimensional
Higgs scalar fields. Then, from the interaction term (35) and the Yang-Mills 
Lagrangian ${1\over 2}\mbox{ tr }(G_{\a\b }G^{\a\b })$, we can acquire the 
contribution
\be
\mbox{ tr }({1\over 2}\tilde{G}_{\m\n }\tilde{G}^{\m\n } -  
\mbox{}^{*}F_{\m\n}\Phi \tilde{G}^{\m\n}), ~~~ 
(\tilde{G}_{\m\n} \equiv G_{\m\n } - F_{\m\n}S )
\ee
which agrees precisely with the corresponding term in (12) if $\alpha  = 1$. This implies that the graviphoton 
current is no longer given by Eq.(32). Nevertheless,  the graviphoton field equation is 
still satisfied for rotationally symmetric fields, thereby yielding dyon solutions.
However, an exact connection between our model with the supergravity model requires 
solving constraints imposed by the dimensional reduction and also a proper consideration 
of other fields in the six-dimensional supergravity theory which do not appear in the 
action (12). This is still an open problem.
Another interesting issue is that of electromagnetic duality in our model. Investigation 
of these issues is expected to shed a new light in the study of 
curved-space monopole/dyon systems. 
 
\vskip .2in
\centerline{\bf ACKNOWLEDGEMENT}
\noindent
We thank Kimyeong Lee and G.W.Gibbons for useful discussions. This work is supported 
in part by the program of Basic Science Research, 
Ministry of Education BSRI-95-2418/ BSRI-95-2442, and by Korea Science and Engineering 
Foundation through the Center for Theoretical Physics, SNU.
 
\vglue .2in

\def\Item{\par\hang\textindent}

{\bf REFERENCES }
\Item {[1]} G.W. Gibbons, D. Kastor, L.A.J. London, P.K. Townsend and J. Traschen, 
Nucl. Phys. {\bf B416} (1994) 850.
\Item {[2]} M.K. Prasad and C.M. Sommerfield, Phys. Rev. Lett. {\bf 35} (1975) 760; 
   E.B. Bogomol'nyi,  Sov.J. Nucl. Phys. {\bf 24} (1976) 861.
\Item {[3]} G.W. Gibbons and C.M. Hull, Phys. Lett. {\bf 109B} (1982) 190.
\Item {[4]} S. Coleman, S. Parke, A. Neveu and C.M. Sommerfield, Phys. Rev. {\bf D 15} (1977) 544.
\Item {[5]} G.W. Gibbons and P.K. Townsend, Phys. Lett. {\bf 356B} (1995) 472.
\Item {[6]} E. Witten and D. Olive, Phys. Lett. {\bf 78B} (1978) 97.
\Item {[7]} A. D'Adda, R. Horseley and P.Di Vecchia, Phys. Lett. {\bf 76B} 
(1978) 298.
\Item {[8]} B.de Witt and A. Van Proeyen, Nucl. Phys. {\bf B245} (1984) 89;
 E. Cremmer, C. Kounnas, A.Van Proeyen, J.P. Derendinger, S. Ferrara, 
B. de Witt and L. Girardello, Nucl. Phys. {\bf B250} (1985) 385. 
\Item {[9]} L.J. Romans, Nucl. Phys. {\bf B269} (1986) 691.

\end{document}